**Defect and Temperature Dependence of Tunneling in InAs/GaSb Heterojunctions**

Ryan M. Iutzi, Eugene A. Fitzgerald

*Department of Materials Science and Engineering, Massachusetts Institute of Technology, Cambridge, Massachusetts, 02139, USA*

*Abstract:* Tunnel field effect transistors (TFETs) utilizing semiconductor heterojunctions have shown promise for low energy logic but presently do not display subthreshold swings steeper than the room-temperature thermal limit of 60 mV/decade. These devices also show a pronounced temperature dependence that is not characteristic of a tunneling process. Herein, we explore these aspects by studying the temperature dependence of two-terminal InAs/GaSb heterojunctions, allowing for the true nature of tunneling at the interface to be seen without convolution from other three-terminal parasitic effects such as gate oxide traps. We compare the temperature dependence of peak current, excess current, and conductance slope for InAs/GaSb interfaces with and without interface defects. We identify that the tunnel and excess currents depend on temperature and defect density and propose that the ultimate leakage current in TFETs based on these materials will be affected by defects and inhomogeneity at the interface. We determine that the conductance slope, a two-terminal analog to subthreshold slope, does not depend on temperature, contrasting sharply with the heavy temperature dependence seen in three terminal devices in literature. We propose that TFETs based on this and similar materials systems are dominated by parasitic effects such as tunneling into oxide trap states, or other parasitics that are not intrinsic to the heterojunction itself, and that in the absence of these effects, the true steepness from band-to-band tunneling is limited by defects and inhomogeneity at the interface.*

Tunnel field effect transistors (TFETs) have become a heavily-explored avenue for the pursuit of low-power electronics due to their predicted ability to obtain subthreshold swings steeper than the thermal limit of 60 mV/decade[1]. Heterojunction TFETs with a small staggered gap (type-II band alignment), or a broken gap (type-III band alignment) offer the potential for higher on-currents due to their small effective band gap, allowing for speeds competitive with CMOS[1]. These heterojunction devices are expected to operate in a regime where the band-to-band tunnel current is limited not by the tunnel-probability, but by the density of states, allowing the subthreshold swing to become limited by the steepness of the 3D band-edges, or 2D band-edges in the case of confinement[2]. However, these devices have not obtained subthreshold slopes steeper than the classical limit[3–9]. Furthermore, temperature-dependent measurements frequently reveal a pronounced temperature dependence that is similar to that of a thermal distribution of carriers with an energy barrier[5,8,9] —an unexpected result given that the band-edges are intended to cut off the Fermi-tails of the tunneling carriers. The cause of both this poorer-than-expected subthreshold swing and temperature dependence remain important problems to be understood and solved.

We have previously demonstrated that the potential for steep subthreshold switching is limited by materials defects via their defect states and the spatially-varying band alignment they cause[10]. We have measured this by fabricating two-terminal diodes and measuring their current-voltage (I-V) characteristics. We are able to extract a two-terminal analogue to the subthreshold swing by analyzing the absolute conductance (G) as a function of voltage ($V_a$), and extracting a "conductance slope" defined as $dV_a/d\log(G)$. This method has been used previously[2,10] and is very useful for predicting subthreshold swings in the absence of three-terminal parasitics which can mask the true effect of tunneling at the interface.

In this work, we study the temperature dependence of two-terminal devices to gain insight into the effect of defects and material inhomogeneity on the operation of TFETs, including their effect on leakage current, and whether or not these devices remain limited by defects and inhomogeneity beyond a certain minimum threshold of material quality, or if they enter a regime where they become limited by other effects, such as phonons. Additionally, we compare our two-terminal temperature dependence of conductance slope to the published three-terminal temperature dependencies of subthreshold swing to provide strong evidence that such TFETs are not switching solely due to band-to-band tunneling as originally expected.

Two structures were grown in this study, both using a Thomas Swan/Aixtron low pressure metalorganic chemical vapor deposition (MOCVD) reactor with a close-coupled showerhead. The first structure was an 80 nm InAs film on a p-GaSb substrate, and the second was an 80 nm GaSb film grown on an n-InAs substrate. Both structures had a 100 nm homoepitaxial layer grown on the substrate. All InAs layers were Si-doped n-type ($1\times10^{17}$ cm$^{-3}$) and all GaSb layers were Si-doped p-type ($1\times10^{17}$ cm$^{-3}$). Both samples had a 20 nm contact layer that was n+ for InAs and p+ for GaSb. All layers were grown at a growth temperature of 530°C and a total pressure of 100 torr, using TMGa, TMIn, TMSb, and AsH$_3$ as precursors and H$_2$ as a carrier gas. Both structures were characterized with transmission electron microscopy (XTEM). Circular mesa diodes were fabricated using electron-beam evaporated Ti/Pt/Au contacts which were used as an etch mask for a self-aligned mesa. A Ti/Pt/Au contact was also evaporated onto the back of the substrate to form a back-contact. Current-voltage (I-V) curves were measured with an Agilent B1500a semiconductor parameter analyzer on a Lakeshore Cryogenic probestation, using liquid nitrogen cooling for temperatures down to 77 K and liquid helium cooling for temperatures down to 4 K.

Figures 1(a) and 1(b) display I-V curves measured at select temperatures for GaSb-on-InAs and InAs-on-GaSb diodes, respectively. The inset shows cross sectional TEM images of the samples, on the 220 diffraction condition, allowing for contrast from crystal defects. As we have reported previously[10], for these reported conditions in MOCVD, InAs films grown on GaSb are heavily defective with many threading dislocations visible in XTEM , while GaSb films grown on InAs are relatively less defective, with no visible threading dislocations in cross section. As temperature is decreased for both devices, there is a visible increase in peak current, an increase in conductance at zero-bias, and a decrease in excess-current beyond the negative differential resistance (NDR) region. An increase in peak current implies an increase in the number of carriers tunneling, and the increase in conductance at zero-bias is also a result of this. However, it can be seen that the conductance through the origin continues to increase even after the peak current has stopped increasing. This is because the conductance also depends on series resistance, and we expect an increase in substrate mobility at lower temperatures. Since this mobility effect is separate from the tunneling, and since the peak current is independent of this effect, we choose to focus on the peak current effect since it provides a clearer look at the tunnel current.

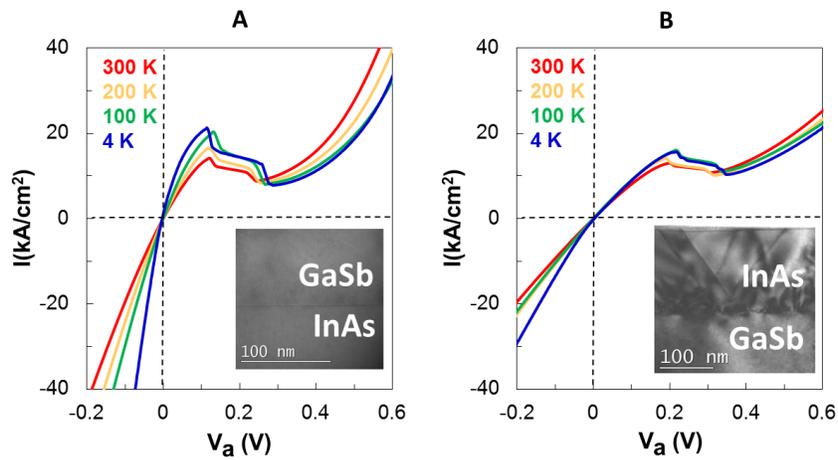

*Figure 1: I-V curves at varying temperatures for (a) GaSb-on-InAs diodes and (b) InAs-on-GaSb diodes. Inset shows a cross-sectional TEM (XTEM) image on the 220 diffraction condition for each sample*

The peak current of both diodes as a function of temperature is shown in figure 2(a) for all temperatures measured. The increase in peak current as temperature is decreased has been reported previously in InAs/GaSb[11] and similar materials systems[12], and is attributed to smaller thermal tails as the temperature drops, allowing a greater percentage of all carriers to have low enough energy to be within the region of band overlap to tunnel. However, as the temperature decreases below about 60 K, we have consistently observed all diodes on both samples show an inversion of the trend, and the peak current begins to decrease with decreasing temperature. Below this temperature, the vast majority of carriers are low enough energy to tunnel, but there are now fewer total carriers due to the fact that the Fermi-level increase with decreasing temperature has resulted in less of an accumulation region near the interface. To illustrate this, Figure 2(b) shows a comparison of the band-edges near the interface at 300 K and 4 K, calculated from a solution of the Poisson and Schrodinger equations, with the temperature-dependence of band gaps included via Varshni parameters. At 300 K, it is likely that there is quantum confinement and a very high density 2D electron and hole gas present at either side of the interface. As the temperature drops, this confinement is lost, and at 4 K, the confinement is minimal and the accumulation region is not deep, which likely leads to the decrease in peak current at these low temperatures.

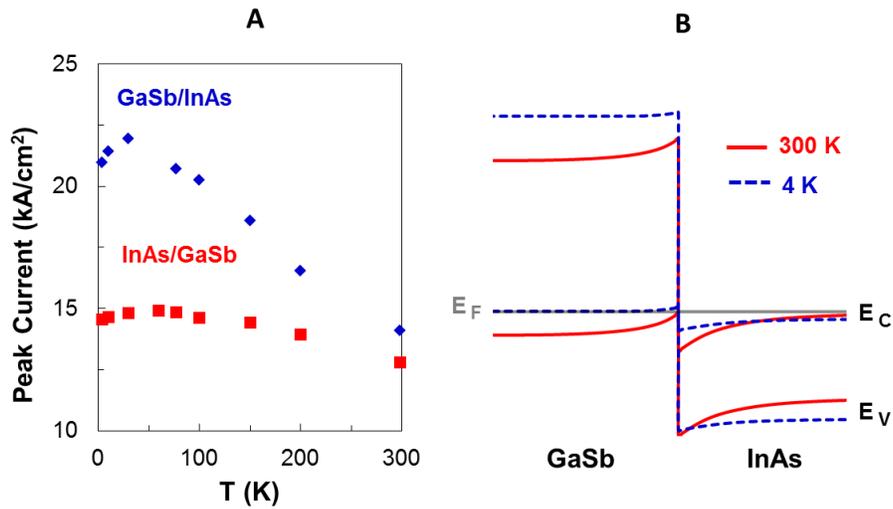

*Figure 2: (a) Peak current vs temperature for both devices, (b) Band diagrams for an InAs/GaSb heterojunction at 300 K and 4K*

Interestingly, the temperature dependence is considerably more pronounced for the higher material quality GaSb/InAs diodes. Again, differences in series-resistance between the two samples can be ruled out as the cause, since series resistance does not affect the peak current. Differences in doping are also unlikely, since both samples are doped identically. Therefore, it is likely that the interface itself has caused this difference, likely due to differences in band-alignment that cause a shift in the fraction of carriers that are originally low-enough energy to tunnel, or by a more pronounced-loss of confinement and accumulation at lower temperatures for the more defective interfaces.

In contrast, when tunneling has stopped, the valley and excess current show a much weaker temperature dependence. Figure 3(a) shows the current of a GaSb/InAs device as a function of inverse temperature at various voltages, beginning near the valley voltage. The temperature dependence is stronger at higher temperature and then reaches a plateau. There is a slight increase in current at higher bias and lower temperature, which we attribute to the large increase in substrate mobility at very low temperatures. The temperature dependence with a plateau is similar to what was reported previously for resonant tunnel diodes in this material system[13], and the trend was attributed to hole current: thermionic hole emission at high temperatures and hole Fowler-Nordheim tunneling at lower temperatures/higher bias, as illustrated in figure 3(b). We observe higher currents and a general shift of the curve's features to higher temperature for our data, and we attribute this to the absence of an AlSb tunnel barrier, which results in Fowler-Nordheim tunneling being favored over thermionic emission up until a higher temperature, due to the higher tunnel probability. However, at low temperature and bias, our currents are still considerably high to be explained by any hole process given the low energy of the hole distribution. Such an anomaly was also observed in the resonant tunnel diode[13] and was hypothesized as potentially being due to dislocations. In figure 3(c), we compare the valley currents of both samples, and see that while the shape of the temperature dependence is essentially the same, the valley current is lower for the dislocation-free diode, and this is consistently observed across all diodes on both samples, and consistent with our previous report[10]. This may be a confirmation that trap-assisted tunneling from the dislocations and other defects (as shown in figure 3(b)) is occurring. The difference in current is quite small[14], and this may be an indication that other defect states are still present in the GaSb/InAs diode.

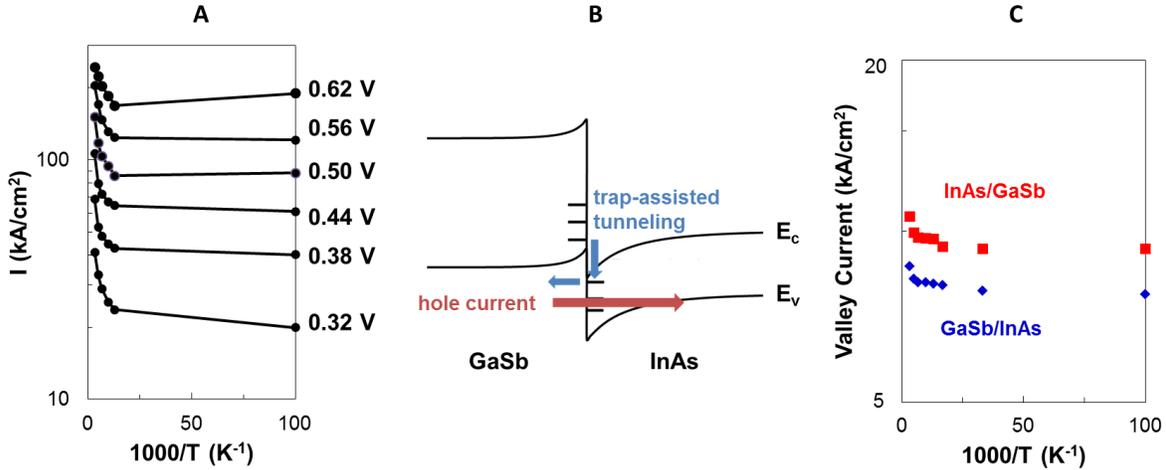

*Figure 3: (a) Temperature dependence of the current at various bias points beyond the NDR region for GaSb/InAs diode. (b) Band diagram for a diode biased at 0.15 V, showing an example of a trap-assisted tunneling leakage route, in addition to a hole current (c) Valley current temperature dependence for both the GaSb/InAs and InAs/GaSb diodes*

Finally, we can extract a conductance slope from both devices as a function of temperature. Figure 4(a) shows the absolute conductance vs voltage curve for the GaSb/InAs device at varying temperatures. Since the NDR region contains oscillations, we have taken an average conductance slope across the instability region, as has been done previously[2,10]. Figure 4(b) shows the conductance slope extracted as a function of temperature. As can be seen, the temperature dependence is extremely weak, and after correcting for the decrease in series resistance as temperature decreases, the temperature dependence essentially disappears. Series resistance was extracted from the series resistance-limited regime of the diode at high forward bias. When we compare to the conductance slope values for the more defective InAs/GaSb device, we see they are less steep, and also have no temperature dependence. We have shown previously that the more defective devices have less steep conductance slope due to more defect states and less-unifiorm band-alignment.[10] We also showed that all devices gained improvement with annealing, which was attributed to a reduction in point defect concentrations and better smoothing of intermixing and point defect concentrations. Such a result indicated that even the dislocation-free samples were still limited by point defects and material inhomogeneity, and this is consistent with the valley current result, where the removal of dislocations gives only a slight decrease. The lack of temperature dependence of conductance slope is consistent with this, as it indicates that the steepness for both devices is still limited by defects and material inhomogeneity, even when the interface appears structurally perfect in XTEM. Such inhomogeneity and defects overpower any thermal-blurring effects such as band-edge blurring due to the deformation potentials caused by phonons, or phonon-assisted tunneling after the bands are misaligned.

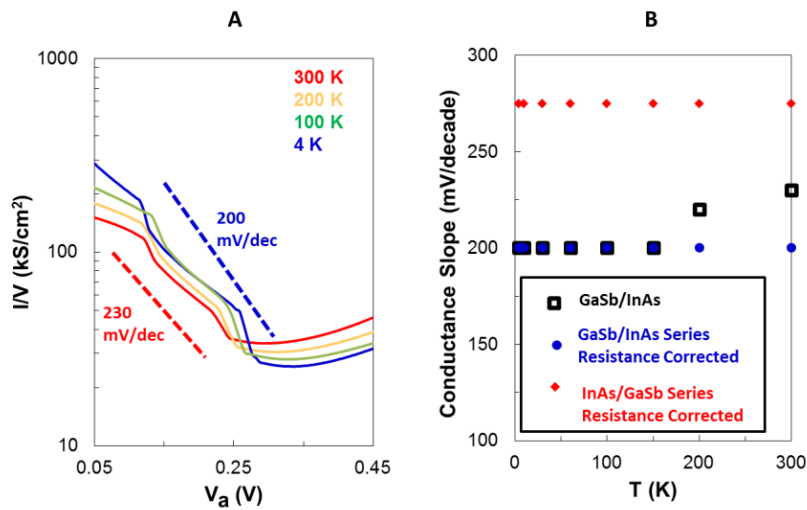

*Figure 4: (a) Absolute conductance-voltage curve at select temperatures for GaSb/InAs diode. (b) Conductance slope as a function of temperature for both diodes. For GaSb/InAs, a comparison of the results before and after correcting for series resistance is displayed.*

The lack of temperature dependence of the conductance slope in two-terminals contrasts sharply with studies that have measured the effect of temperature on subthreshold slope in three-terminal TFETS of similar materials systems including InAs/GaSb[5], InGaAs/InP[9], and InGaAs[8]. Mookerja[8] proposed that in the region of steep-subthreshold switching, current is not true band-to-band tunneling, but tunneling into trap states at the gate-oxide/semiconductor interface and subsequent generation into the conduction band. Our results support this hypothesis, since two terminal measurements were chosen specifically to avoid a gate oxide, and in the absence of it, there is no temperature dependence. Given that published heterojunction TFETs show a temperature dependence of subthreshold slope, the fact that our two-terminal measurements do not is a strong indication that the subthreshold swing in most heterojunction TFET devices is not caused by band-to-band tunneling, but by trap-assisted tunneling to gate-oxide states and subsequent generation, or by some other mechanism that is specific to three-terminal devices and not intrinsic to the heterojunction itself.

In conclusion, we have measured two-terminal I-V curves on devices with and without high defect densities. We identified that the band-to-band tunneling current temperature dependence changes with defect density, and that once the bands are misaligned, the excess current is likely due to hole current at higher temperatures, and may likely be due to trap-assisted tunneling at lower temperatures/lower biases. We have determined that the conductance slope does not have a temperature dependence. This indicates two important conclusions about TFETS: 1) The subthreshold swings in many heterojunction TFET devices are affected by three-terminal parasitics that are not a true function of the band-edges, and 2) In the absence of three-terminal parasitics, the subthreshold swing that could be obtained is likely limited by defects and material inhomogeneity at the heterojunction interface, even in the case of what appear to be structurally perfect interfaces in XTEM.

The work was supported by the Center for Energy Efficient Electronics Science (NSF Award 0939514).